# On Some Recent Insights in Integral Biomathics

Plamen L. Simeonov
JSRC
Berlin, Germany
Email: plamen@simeio.org

**Abstract**: This paper summarizes the results in Integral Biomathics obtained to this moment and provides an outlook for future research in the field.

# 1. What are the challenges in formalizing biology today?

Formalization of any kind presumes the prior presence of something concrete and particular. Recognizing this in biology is not easy. Many biological phenomena do not have adequate mathematical representations. This is because living systems are deploying logic and semiotics beyond our conception of mathematics into the domain of computation, which is on its part much richer than the standard Turing machine paradigm. In particular, there are four areas deserving attention with broad impacts in and beyond biology (Root-Bernstein, 2012a):

1. ***A theory of self-emergent objects*** that can carry out functions within interactive variances of the constituents of living systems: developing models of self-emergent objects (origins of first cells; self-assembly of viruses, etc.) that carry out functions (selecting/rejecting among possible components; minimizing free energy; etc.) while utilizing both continuous and discrete information. Set theory is too abstract to handle such "objects"; they have more structure and other properties than sets and the elements of sets. Therefore self-emergent objects must be defined in a frame where their different qualities can be described. Such a theory should be able to incorporate the work that has been done on understanding hierarchical systems, emergent properties, complexity theory, etc. Its mathematics should therefore be extraordinarily integrative. A subsumptive or specification hierarchy is more general than set theory and could be the first step in this direction.

   There are two fundamental issues to be taken in consideration. First, cells are autonomous/autopoietic – they form themselves. Therefore, we need to use a frame in which we can describe self-organizing / self-emergent objects mimicking "cellular life" as evolution of such objects using specifications of the relations between them that do provide rules i) to limit the entry and exit of individual elements, and ii) allow elements to undergo transformations (e.g. metabolism) within the object. A key question to answer here is wether the *self* (the individuum with a distinct identity) is a prerequisite for or a consequence of the self-organization? Second, biological objects (cells) have the variance property[1]. We need a theory that allows the definition of objects that are not characterized by specific numbers, proportions and rates of turnover, excretion and replenishment of the object/cell constituents, but i) by (empirically determined) *variances* (number ranges) within which all of these constituents must exist in order for the living object to function as such (Zadeh, 2000; Kauffman, 2001), and ii) with the *self* being *potentially* associated with these variances.

   This "bio-affine" theory must allow the integration of continuous and "grainy"/discrete temporalities for the same "object" simultaneously: a circular relation of i) handling continuous variations of the chemical kinetics, e.g. continuous interactions/flows of elements, while ii) acting on small sets of definable/discrete elements/individuals (calculations of modular probabilities) determined by the chemical kinetics. Also, it should have operators/functions capable of defining *state-sensitive objects*, i.e. to model switching processes between stable states when certain values or variances (within them) are exceeded.

2. ***A theory of complementary assembling***: biological systems build and organize themselves based on the principle of molecular complementarity to produce robust aggregates/modules. Such a theory is therefore important, because the formation of complementary (symmetric or asymmetric/chirality) modules within complex systems can prune out huge numbers of unfeasible possibilities at each step of the hierarchical modular assembly.

---

[1] Any given cell must have chromosomes, but their number can vary (as they do in cancers and parthenogenesis) and still be viable; they can have many ribosomes and mitochondria or but a few and still live; they can accumulate certain amounts of toxins or lose a certain amount of key ions and still function; etc.

3. **A *qualia jump*[2] theory**: Mathematics generally treats either scalar quantities or vector quantities (statically defined), but not the transformation of scalar to vector (dynamic type transitions). However, some properties of biological systems involve transformations from pure scalar to pure vector quantities (and vice versa): e.g. a chemical neurotransmitter signal (scalar diffusion) becomes a directional electrical signal (a vector).

4. **A hidden morphology theory**: the linkage between form and function.

    The mathematical challenges involved in attempting to model biological form-function interactions are far from trivial. Natural selection attempts to optimize forms to carry out particular functions, but since novel functions evolve from existing forms, these formal attempts may be seriously limited. On the one hand, we do not have geometrical tools that can easily model processes such as the complex folding of proteins or chromosomes, or detailed embryological development. Mathematical forms share little with the actual biological processes that give rise to these structures. The mathematical abstractions currently used in system biological models generally do not illuminate the processes that give rise to biological geometries, but only their outward forms despite the work of René Thom (Thom, 1994). What is interesting about biological forms, however, is not their geometry per se, but the ways in which these forms are *reifications of the biochemical processes* which they carry out or make possible. For example, it has become evident that the folding of chromosomes is a prerequisite to bringing together genes that would otherwise be spatially separated; and also that spatial proximity permits the rapid diffusion and control of interactive gene products that would otherwise be unable to interact in a reasonable biological time frame across an unfolded genome (Junier et al., 2010). Similarly, in human developmental biology we have now excellent data concerning the sets of genes that must be turned on and when they must be activated or inactivated in order to produce proper embryological development, yet the *discrete information* generated from combinations of individual genes is expressed as a *continuous flow* of proteins and hormones that produce gradients which must be reified as organized groupings of cells that have a specific form. So embryology is also stymied by the lack of mathematical approaches that can link discrete, continuous and geometrical information simultaneously. But what kind of mathematical notions would make it possible to model simultaneously the effects of geometry (spatial structure) on continuous functions such as diffusion that in turn regulate on-off gene regulatory switches that act discontinuously or digitally? And how can a mathematical object obtain and maintain its identity? Also, Lewontin has stressed the *reciprocal relationships between genes, organisms and their environment*, in which all three elements act as both causes and effects (Lewontin, 2002). Therefore, we should be able to explore alternative avenues to traditional unidirectional genotype-phenotype relationships such as cyclic or helical or chaotic genotype-phenotype mappings.

## 2. Are there any solutions in sight?

In fact, such biologically relevant mathematical concepts providing the base for realizing the scenario in the previous section *do* currently exist. The above questions reflecting work by (Root-Bernstein, 2012a) were raised with reference to a '*classical*' view on the different domains of mathematics. This traditional view on mathematics has been modified with the introduction of more unified and 'relational' approaches, such as category theory (Mac Lane, 1998) and algebraic geometry (Felix et al., 2008) in general, as noted by (Hoffman, 2013). In addition, richer Bayesian, contextual models, but also quantum models (Gomez-Ramirez &

---

[2] A quale is an individual instance of subjective, conscious experience, "an unfamiliar term for something that could not be more familiar to each of us: the ways things seem to us" (Dennet, 1988).

Sanz, 2013; Kitto & Kortschak, 2013; Gabora et al.; 2013), and post-Turing computation (Siegelmann, 2013) and elaborate forms of logic (Goranson & Cardier, 2013; Marchal, 2013) have changed the understanding about the role of mathematics in biology. The following two sections recapitulate two complementary approaches which both touch on the implementation of the Integral Biomathics strategy of (Simeonov et al., 2012b).

## 2.1 Category Theory (CT), Memory Evolutive Systems and Relational Biology

Category Theory (CT) is a *formal domain* of mathematics at the frontier between mathematics, logic and meta-mathematics. It provides conceptual means for thinking about mathematics in terms of the relations between its domains. CT is: i) a language which tries to uncover and unify the operations made in different mathematical branches (such as abstract algebra, topology, geometry, etc.); (ii) a philosophy, which provides a general notion of mathematical structure (e.g. category of sets, of groups, of graphs, of fields, of rings, etc.); and iii) a *mindset* providing new tools (such as adjoint functors, colimits, etc.) that are useful for defining models of complex systems.

Whereas the question 'Can we model this with differential equations?' is a valid one with a yes/no answer, the answer to 'Can CT provide tools to handle this?' is often "yes as long as" we phrase the particular category-theoretic descriptions appropriately. But this affirmative answer is due to the generality of CT and means also that it is not sufficient as such for modeling things 'concretely'. One must still use a specifically suited subbranch of mathematics, although casting the problem directly in the language of category theory is also often useful. Category Theory offers a 'relational mathematics' emphasizing the relations between objects rather than the objects themselves. Indeed an object of a category is characterized, not by its 'ontological structure' but by the set of morphisms arriving at it (Yoneda Lemma), which corresponds to the different operations in which the object participates. Therefore there is no need for a new kind of set theory (Question 1), but a whole variety of morphisms can be used instead of simple maps, thus giving one much more representational freedom. For instance, we may take a category whose objects are molecules, and the morphisms model chemical interactions between them. We can look at a molecule as the set of its atoms and their chemical bonds. However, the morphisms between molecules do not correspond to maps between these sets but represent emergent properties at the molecular level depending on quantum level properties with infinitely many elements.

Two category-theoretical approaches are of particular interest for us: Memory Evolutive Systems, MES (Ehresmann & Vandbremeersh, 2007; Ehresmann, 2012) and Relational Biology, RB (Louie, 2009, 2013), with the latter forming a continuation of Robert Rosen's work on the subject (Rosen, 1958a/b; 1959; 1991; 1999) and on anticipatory systems (Rosen, 2012). Both represent domain-specific incarnations of CT for biology as synergetic syntheses of a number of mathematical theories (partially ordered sets, lattices, graphs, categories, adjacency matrices, and interacting entailment-type networks, etc.).

More specifically, the MES theory can handle the different problems with self-emergent objects in the following way:

1. A "self-emergent object" is modeled as a *colimit* emerging via a complexification of an appropriate category. Examples are given in (Ehresmann & Vanbremeersch, 2007).

2. MES are particular state-transition systems. A component of the system (for instance, a cell) is not represented by a unique object but by the family of its (infinitely many) successive states. Indeed, in an MES, the original system is not modeled by a unique category but by an evolutive system, that is, a family of categories indexed by time with transition partial functors applying between them. Thus, a component C can be modeled using the family of its successive states. This representation allows one to differentiate between the complex identity (or "class identity") of C and the individual identities of its lower level components.

Second, complementary assembling (Question 2) is allowed through the consideration of "activator links" in memory from a molecule to the procedure which consists in attracting the complementary molecule.

The qualia jump theory (Question 3) is related to hierarchy and emergence, whereby the Multiplicity Principle plays an important role since it allows for the emergence of new properties (represented by complex links). For instance, the electric signal at the cell level is an emergent property from a pattern of chemical signals acting on the molecular level.

The "hidden form" problem (Question 4) refers to the emergence of a higher-layer object as the colimit of a pattern of interacting lower-layer objects (which forces its geometric form). The key question remaining in all cases is how to implement the colimits empirically.

The RB approach presented in *The Reflection of Life* (Louie, 2013), a domain-specific CT for biology is formulated in terms of *set-valued mappings*. Instead of considering a mapping of the form $f : A \to B$ (from set $A$ to set $B$, sending element $a \in A$ to element $f(a) \in B$), Louie suggests the new correspondence $F : A \mathrel{-\ell} D$ or[3] $F : A \mathrel{-[} D$ (from set $A$ to subsets of $D$, sending an element $a \in A$ to set $F(a) \subset D$); particularly consequential hereby is the case when $D = H(B,C)$ (the hom-set $H(B,C)$, a set of mappings from set $B$ to set $C$). By using the theory of set-valued mappings it is possible to handle the above questioned four issues about living systems (1 to 4), of the point/set hierarchy, of continuity/discreteness, and of cyclic/helical entailment with much naturalness and ease. Ultimately, the theory of set-valued mappings culminates in the so-called imminence mapping, which greatly facilitates the further investigation of functional entailment in complex relational networks. In short, both MES and RB provide truly integrative, both multi-scale, multi-temporality mathematical approaches to the modeling of biological phenomena. Their common problem remains that they need to be translated into a more understandable and computable language, that is, an "operational semantics" to handle the implementation of their concepts is needed.

## 2.2 The Wandering Logic Intelligence (WLI)

A recent study presented in the 2012 Oxford seminar on the conceptual foundations of system biology (Bard et al., 2013) has put forward two essential principles of modern biology:

1. ***Feedback between the participants at several levels* – with *oscillations being inherent* in the dynamics of this feedback in order to guarantee stability**.
   This rule defines the "first law of system biology": Any complex biological event involves activity at many levels and the properties that emerge from this activity are not necessarily predictable (Kolodkin et al., 2012).

2. **No *level of preferred status exists,* with necessary and important events taking place at and between each level so that causality occurs upwards, downwards and within levels** (Noble, 2012; 2013). This rule has been shown to be realized within the entire physiome (genome, proteome, interactome) (Werner, 2005; Shapiro, 2013). An open question here is how we can address *supervenience*[4] (Kim, 1984; 1987).

---

[3] The 'forked arrow' notation '$-\ell$' or '$-[$' is Louie's invention.

[4] The term "supervenience" was suggested by Kim to denote such phenomena as moral properties that supervene on natural properties, or mental characteristics that supervene on physical characteristics (e.g. the properties of our nervous system). The term can be defined as follows: For two sets of properties, A (the *super*venient set) and B (the *sub*venient set or supervenient *base*), A supervenes on B just in case there can be no difference in A without a difference in B. Reversing this principle delivers the converse concept of antisupervenience or *determination*: B determines A just in case sameness with respect to B implies sameness with respect to A. Hence supervenience and determination are two sides of the same coin.

These two principles and two additional ones, related to the emergence of the self and the complementarity requirements presented in section 1, are epitomized by the Wandering Logic Intelligence (WLI) theory (Simeonov, 2002a/b/c):

1. Multidimensional Feedback
2. Pulsating Metamorphosis
3. Self-Reference
4. Dualistic Congruence

They were successfully tested in practice for the specification and implementation of mobile computing and communication architectures (Simeonov, 2002c, Wepiwe & Simeonov, 2005). These principles can be used also for the investigation of interacting living systems (Louie, 2013) from both the internalist and the externalist viewpoint, i.e., the $1^{st}$ and $3^{rd}$ person perspectives (Rössler, 1987, 1998; Matsuno, 1989, 1996). This can complement the categorical formalization of MES with the computational semantics of WLI whereby both are mutually reflective.

Into this context fall two recent insights by (Matsuno, 2013) and (Salthe, 2013). The first is about the necessity to elaborate and include a new kind of active temporal logic, which internalizes the description of the metabolic dynamics of a living system recursively *into t*he dynamics itself. This temporal dynamics[5] of the organism (or machine[6]) maintaining its own identity (from a $1^{st}$ person perspective) can be modeled in terms of a *constant updating* of the present perfect tense into the present progressive tense. The second constructive aspect of this computational dynamics and operational semantics type description comes from the macroscopic level ($3^{rd}$ person perspective) and requires one to trace it both spatially, within sets of nested hierarchies, and temporally across the entire development cycle.

The synergies between MES and WLI were already studied in (Ehresmann & Simeonov, 2012). Both theories taken together appear to provide a sound base for the modeling of biological systems.

## 3. What are the foundations we build on?

Integral Biomathics strongly endorses the following topics for research in biology:

1. A source of current methodological problems in system and molecular biology lies in a preoccupation with the quantitative methods and a *disregard for the qualitative, structural approach* (Schroeder, 2013).
2. The problem of *combinatorial complexity*[7] when dealing with large dynamic multi-protein complexes in living systems leads to the insight that conventional modeling approaches (like differential equations) fail to describe the self-assembly process, mainly due to the combinatorial explosion of the number of intermediate complexes (Tschernyschkow et al., 2013). A hierarchical levels approach could be a solution.
3. *Category theory* is a powerful means to express and discover new syntactic properties different from all others currently considered in biology (Cazalis, 2013).
4. *Algebraic geometry shows that form follows function.* In order to be properly understood, this requires one to take into account the interaction between form and function. The expression "G × M → M" encapsulates this relations between structure and function[8]. It is the task of biologists to provide the particular structure of the parameter group G that is involved in a biological phenomenon. In regard to the traditional distinction between *pure* mathematics and applied mathematics, G in the expression G × M → M, G can be anything (from graphs to categories), and M can be anything from a module to a category. In view of the tremendous variation that biology offers, choices must be made in the application of mathematics to particular biological phenomena. Hereby algebraic geometry or 'rational homotopy theory' can prove to be essential to understanding the structure-function map in biology (Hoffman, 2012-2013; Felix et al., 2008).

---

[5] expressed through the constant exchange of its constituent molecules by new ones from its environment
[6] in the post-Gödelian definition of (Marchal, 2013)
[7] Tschernyschkow et al. show how to apply a novel rule-based modeling approach in space to the study of the formation of the inner kinetochore structure that plays a central role in chromosome segregation and cell division. The pertinent simulation experiments require unconventional mathematical and computational methods for their interpretation. Structure clustering, information theory, phylogeny, and visualization have been applied so far. The challenges laid open in the present article document the need for new mathematical and computational approaches in order to tackle the combinatorial complexity of living systems in space.
[8] Often biological form is changing during actions.

5. *Biocomputation and temporality*: It turns out that computation as now performed is unnatural in biology. Computations conceived in first-order logic are generally *not decidable*. This fact implies that there is no guarantee that such an algorithm can lead to an effective procedure to decide upon membership in a legitimate domain of discourse, or on an allowed set of formulas using Boolean true or false values[9].

6. *Development and decidability*: A crucial question here is whether and how *decidability* is implemented in the developmental process of biology. The current practice of employing simulations in biology without examining the possibility of undecidability is a bit ill conceived. What makes a developmental process maximally challenging from the point of view of computation is whether or not decidability[10] can be obtained (Salthe, 2013).

7. *Use of probabilities:* The limitations of traditional or standard mathematical statistical models including the standard Bayesian approach, for modeling biological systems require the revision in terms of the *inverse* problem to models and parameter values being obtained by measurements. Here new methods such as a truly Bayesian approach (Gomez-Ramirez & Sanz, 2013) are suggested as effective tools to deal with biological probabilities e.g. in neuroscience[11] and physics.

8. *Realism and fractalization*: A resolution of the dichotomy between the one-person and third-person description of systems requires a subtle account of the interaction between multiple-time scales to rigorously describe complex internal (first person) experiences modeled externally (from a third person's perspective) as hierarchical nested events (Vrobel, 2013).

9. *Objectivism and contextuality*: There is a general problem in using the Newtonian Paradigm in the study of life. Specifically there is the difficulty of defining *units of selection* (Salthe, 2008) in biology[12], where this concept plays a fundamental role. This is an unavoidable consequence of the need to take into account the *context* of whatever biological process is considered. This applies already on the molecular level when interaction between parts of the genome are considered, and on a larger scale when genotype-phenotype interactions are important. A reason for the failure of elements of the Newtonian Paradigm applied in the study of life are the two concepts of an *object* and *objectivity*. This problem was proposed to be fixed by introducing a kind of contextual mathematics using three aspects of Quantum Field Theory (QFT): potentiality in terms of multiple stable lowest energy states, spontaneous symmetry breaking and Nambu-Goldstone emergence, to explain the creation of new biological objects and structures (Kitto and Kortschak, 2013). However, a tough problem lurking here is how to secure the right fundamental predicates for the intended enterprise.

10. *Quantum-like potentiality in evolution*: A new mathematical framework for preadaptation that defines the state of a trait as a linear superposition of basic states – or possible forms – is presented here as mutually orthogonal weighted eigenvectors in a complex Hilbert space, into which a trait could evolve. The choice of trait changes is expressed as an adaptive quantum function, which plays the role of the observable (Gabora, Scott and Kauffman, 2013).

The realization of quantum-like potentiality in evolution (Gabora et al., 2013) in combination with other seminal ideas like the contextuality concept (Kitto and Kortschak, 2013), the framework of the cognizer-system model (Nakajima, 2013) or the full Bayesian approach (Gomez-Ramirez & Sanz, 2013), or the formal categorization of the individual (Catzalis, 2013), or the form-follows-function mapping of algebraic geometry (Hoffman, 2013) and the third person expression of the internal fractalization of time (Vrobel, 2013), are all consistent with the line of thought developed in the INBIOSA project (Simeonov et al., 2012a/b). These approaches are far from being congruent but we believe that they provide a much-needed new focus that predictably entails a significant "cultural exaptation".

---

[9] This is the satisfiability problem (SAT) within the frame of non-deterministic Turing-machine computation, which is crucial with regard to the likelihood of P vs. NP completeness (Cook, 1971).

[10] either from formal logic ,or from bare empirical facts by way of an act of measurement
[11] neuroscience has the 'coarse to fine' trope (making a subsumptive hierarchy)
[12] This issue is about the *nature of identity* of a material body encountered in biology.

## 4. What are the potential application domains?

The above enumeration of areas of application addressed by Integral Biomathics techniques is incomplete. For example, the WLIMES approach (Ehresmann & Simeonov, 2012) is able to develop a robust computational framework on top of a qualitative formal model of living systems based on an evolutionary category theory in the domain of personalized medicine addressing such fields as:
- *virology/vaccinology*
- *molecular genetics*
- *oncology*
- *neuroscience*
- *immunology*
- *emergency medicine*

Characteristic of these domains is the multi-layered nature of the phenomena. Key features of these domains are that they are data and knowledge intensive fields, and beset by incomplete and noisy information.

Infection biology and in particular *neurovirology* combine a number of key systemic themes in the above 6 areas: the interactions between organisms and their environments, and those between two living systems. One of the challenging questions that can help better prevent spreading of infections and pandemic diseases is understanding how viruses move within hosts, e.g. from the skin to the nervous system (Zaichick et al., 2013) and thereby attack multiple weak points in an organism's defense system.

## 5. How are these challenges being approached today?

Today, scientists use a multiplicity of mathematical and computational approaches based on the classical line of Turing machine models. None of the challenges facing the formalization of living systems outlined in section 1 have been addressed so far (Simeonov et al., 2011).

There are two main fields for theoretical research in biology and medicine:

- quantitative modeling: ODE/PDE systems, probabilistic/statistical/stochastic (e.g. Monte Carlo, Markovian, Bayesian etc.) models

- qualitative modeling: discrete state-transition systems such as Boolean/logical networks, Petri nets, process algebras (CCS, CSP, ACP, LOTOS), etc.

A third field is concerned with the visualization of experimental/empirical evidence:
- regulatory charts/maps/graphs, 3D VR models, simulation and animation, etc.

The combination of these areas is known as "hybrid modeling". Basically we distinguish between the hypothesis-driven (qualitative) and the data-driven (quantitative) approaches, both using visualization methods to validate their models, yet with (almost) no links between them. The funds spent for carrying out data-driven research (Human Genome Project, Human Brain Project, etc.) exceed by orders of magnitude those for exploring hypothesis-driven research.

### 5.1 Qualitative vs. quantitative models

An important current limitation of quantitative methods is the deficit in detailed quantitative information about the reaction kinetics (kinetic and their associated parameters) underlying the protein–protein and protein–DNA interactions (Alm & Arkin, 2003).

The missing data are usually replaced by the use of probabilistic and statistical methods. For instance, (Wierling et al., 2012) propose to predict the disturbances induced by targeted cancer therapies on the EGF signaling network by introducing a Monte Carlo type strategy incorporating repeated simulations with parameter vectors sampled from an assumed random distribution with subsequent statistical significance testing. They demonstrate the applicability of this approach by generating statistically reproducible predictions of potential drug targets.

Since the application of standard statistical models in biology is limited, refinements of the heuristics are suggested to fill this gap (Nakajima, 2013; Gomez-Ramirez & Sanz, 2013). On the other hand, the predictive power of qualitative approaches like Boolean networks and other discrete techniques is limited to qualitative conclusions only.

The majority of work done up until now has been on modeling complex molecular networks. A large body of work was done in quantitative modeling using systems of differential equations (Olmholt, 2013) and in qualitative approaches based on discrete state-transition systems (e.g. by using Petri nets (Heiner & Gilbert, 2013)). However, progress in these areas was limited. This is partly because we know little about the reaction constants and partly because the nature of most of the interactions between the participating proteins has yet to be elucidated qualitatively or quantitatively. We cannot even be sure, for example, that the law of mass action holds true given the small number of proteins within a cell so that models require stochastic or "negotiation" elements.

### 5.2 Consequences

The data driven approach in biology has not held its promise of predictive diagnostics and personalized medical treatment (Ahn et al., 2006; Gomez-Ramirez & Wu, 2012). The extensive knowledge of the genome sequences of human beings and various pathogenic agents has led to the identification of but a limited number of new drug targets (Drews, 2003). The employment of new methods for drug discovery based on strategies like high-throughput screening, combinatorial chemistry, genomics, proteomics and bioinformatics does not currently bring forth the expected new medications and therapies (Kubinyi, 2003; Glassman & Sun, 2004). A number of biotechnological projects such as gene therapy, stem-cell research, antisense technology and cancer vaccination have failed to deliver the expected results (Glassman & Sun, 2004). The common problem with many of these innovative techniques is that the risks and unwelcome side effects have been underestimated, as was the case for gene therapy (Williams & Baum, 2003). Therefore we presently give preference to the qualitative, hypothesis-driven approach, considering science to be not only concerned with producing predictive model but also and more prominently to be involved in *understanding* Nature, as René Thom argued in defense of his notion of qualitative mathematics (Thom, 1977; 1994).

## 6. Why can we do better than before?

Current approaches to complex problems rely on modeling but one aspect of the problem with but one form of mathematics, switching from one to another sort of mathematics to address the next aspect and then to a third one to describe yet another. All this switching is an indication of how inadequate our mathematical tools (ODE/PDE systems, stochastic models, discrete state-transition systems, etc.) of to date are. Biological systems function at all of these levels simultaneously, so why cannot our mathematics do the same? It is not biology that is too messy to be modeled, but our mathematics is inadequate because we are not creative enough (Hong, 2013) to adequately address these biological problems.

**This shows we need to go beyond the standard paradigm of computation as formally implemented in the Turing machine. One candidate to help us get out of the current stifling situation is to pay more attention to the Turing oracle machine[13]. This is a formal scheme to address the issue of making decisions that would consistently survive on the most concrete and particular level in our empirical world. Note that the source data of biology come from concrete particulars without exception.**

Remember that hierarchy theory (Salthe, 2012) suggests that reductionism can never explain how novel properties and processes emerge. Biological entities have properties that differ from chemical and physical ones and require developing appropriate mathematical notions. What we need is not more detailed physical models of biological systems that could handle larger and larger amounts of detailed data from increasingly fine-grained studies of the components of biological systems, but ways of identifying the biological properties that are as unique to such complex aggregations as temperature is to a set of molecules (Root-Bernstein, 2012a). This is why we need a new biomathematics – a mathematics based on such fields as category theory and algebraic geometry which treats continuous functions, sets, vectors, fields and other formalisms in a single complementary framework that is truly *integrative:* A unique evolutionary mathematical and computation platform that deals with the emergence of organization from non-random selection amongst replicating variations within complex populations of things. The challenge of Integral Biomathics is to provide the mathematical and computational methods necessary for the modeling of such emergent properties and organization.

## 7. What is missing in current theoretical biology?

Besides the often-referred to self-organization principles (Camazine et al., 2003), biological systems exhibit the following special characteristics which have been neglected by virtually all qualitative and quantitative, approaches so far:

- *Emergence*: Emergent properties differ from resultant properties that can be predicted from local lower-layer information. They resist premature attempts at being predicted or deduced by explicit calculation. Thus, the behavior of biological systems cannot be understood or predicted by simply analyzing the structure of their constituents. The latter interact in many different ways including multiple negative feedback and feed-forward controls, which lead to dynamical properties that cannot be adequately predicted by using linear mathematical models that disregard cooperation, competition, and non-additive effects. Therefore, in recognition of the complexity of informational pathways and networks, non-classical mathematical formalisms are necessary for modeling such systems (Aderem & Smith, 2004; Root-Bernstein, 2012a). [In our project we are going to model emergence both in an explicit and in an implicit way, i.e. (a) by developing new formalisms and (b) by using simulations, without any explicit reference to the emerging properties[14].] Emergence theory regards the natural world as organized in terms of hierarchies that have evolved over time (Pattee, 1973; Salthe, 1985; Salthe & Matsuno, 1995; Kim 1999, Morowitz, 2002). Whereas reductionists defend an 'upward causation' by which molecular states bring about the higher-layer phenomena, the proponents of emergence advocate a 'downward causation' by which the higher-layer systems influence lower-layer ones (Kim, 1999). We adopt the "inside-out" causation concept of Sydney Brenner (Brenner, 2010) in the context it is represented by Denis Noble (Noble, 2008; 2012). The only way to anticipate emergence is to observe and trace the pattern formations and their relations *within their own context at possibly multiple layers* below, at and above the layer of their initial emergence.

---

[13] An oracle is a machine which computes a single arbitrary (non-recursive) function from naturals to naturals (Turing, 1939). In other words, this is just another name for non-trivial meta-level heuristics that lies outside an object-level theory. In Integral Biomathics, we regard "oracles" to truly lie beyond the object-level (scientific and/or mathematical) theories, such as group theory and QM. In other words, an oracle is anything that *is* or *can lead to* a true statement that cannot be reached within a formalized (syntactic) system of the pertinent theory. Oracles are part of all human knowledge that cannot be proven within any of the *currently known* formal systems; i.e. they contain "true" statements that cannot be proven in a Gödelian sense. All our theories are bound to remain incomplete, but as they become richer and richer, what once lied outside a given theory will become part of the (again still incomplete) new theory.

[14] For instance the shape of a galaxy can be derived by simulation (most models apply Newton's law or variants thereof) over a large number of "particle" stars. The shape of a galaxy then is implicit in Newton's law.

- *Robustness*: Biological systems tend to be impervious to less than drastic changes in their environment because they are able to adapt and have *flexibly redundant* components and pathways that can act as backups if individual components or paths fail (Csete & Doyle, 2002; Kitano, 2002; Ehresmann & Vanbremeersh, 2007).

- *Hierarchical Modularity*: Biological systems are hierarchically organized with subsystems being physically and functionally isolated in such a way that a failure in one module does not spread to all other parts of the structure with possibly lethal consequences (Alm & Arkin, 2003). This modularity, however, does not prevent different compartments from communicating with each other (Weng *et al.*, 1999).

- *Organisational Closure and Openness*: Biological systems are self-contained with internally coupled pathways, but exchange matter, energy and information with their environment and are therefore not in thermodynamic equilibrium (Yoshiteru, 2009; Mossio & Moreno, 2010).

- *Contextuality*: Biological systems are extremely complex and dependent on their local contexts and their embedding within other systems and ecologies (Aerts & Gabora, 2005; Cardinale & Arkin, 2012; Kitto & Kortschak, 2013).

- *Complementarity*: There is a chemical link between biological structure and function ("structure follows function"). First, molecules that bind to each other almost always alter each other's physiological effects; and conversely, molecules that have antagonistic or synergistic physiological effects almost always bind to each other. Second, contemporary biological systems contain an embedded *molecular paleontology* based on small, molecularly complementary modules/subunits that are built into contemporary macromolecular structures such as receptors and transporters. Third, complementary modules are conserved and repurposed at every stage of evolution. Finally, modularity based on molecular complementarity produces a means for storing and replicating information. Linear replicating molecules such as DNA or RNA are not required to transmit information from one generation of compounds to the next: compositional replication is as ubiquitous in living systems as genetic replication and is equally important to their functioning. Chemical systems composed of complementary modules mediate this compositional replication and gave rise to linear replication schemes. Molecular complementarity plays a critical role in the evolution of chemical systems and resolves a significant number of outstanding problems in the emergence of complex systems. Most physical and mathematical models of organization within complex systems rely upon nonrandom/deterministic linkages between components. Molecular complementarity provides a naturally occurring nonrandom linker. More importantly, the formation of hierarchically organized stable modules vastly improves the probability of achieving self-organization. In addition, molecular complementarity provides a mechanism by which hierarchically organized stable modules can form. Complementarity is ubiquitous in living systems, because it provides the physicochemical basis for modular, hierarchical ordering and replication necessary for the evolution of the chemical systems upon which life is based (Root-Bernstein & Dillon, 1997; Root-Bernstein, 2012b). Recently, (Kolodkin et al., 2012) argued for replacing the tradition of Occam's razor in science (reflecting the intention to model the system as simply as possible) with a 'law of completeness'.

While already taking into account hierarchical organization and modularity, the reductionist agendas of molecular and systems biology in the past ignored such essential characteristics of living systems as emergence, robustness, complementarity and contextuality. This restriction had a profound impact on biological and biomedical research during the past 50 years.

For instance, the flexible redundancy (cf. robustness) is at the base of the MES' Multiplicity Principle (MP). It means, so in Edelman's "degeneracy" (Edelman & Gally, 2001) that the same output can be generated by structurally and functionally different components and pathways. The latter are *hidden (potential)* for the researcher a priory due to the switching off of a particular known component or path (e.g. an enzyme that triggers the synthesis of an undesired protein). This principle is a new key insight for understanding such phenomena as cancerogenesis and MODS [15]. It needs thorough consideration when developing future personalized medical therapies. In this way there exists a functional interdependence, not only between robustness and complementarity but also among the other characteristics of living systems. This requires a higher layer of functional modeling addressing the above issues.

In most theoretical biology models the following holds true:
- data-driven and hypothesis-driven models are decoupled,
- all space-scales of interest are non-integrated,
- either upward or downward causation is used alone (generally upward),
- no self-* properties are addressed.

Therefore, current models are incomplete; are sensitive to noisy or missing data; and lack expressiveness. In addition, the modeling environments have no embedded intelligence. As a result, the experimenter/researcher is confronted with an *exponentially growing complexity* that results from a huge number of variables and parameters to explore. In addition, he/she faces *incomplete and noisy data*. The ease of data collection has exacerbated the need of models rich enough to explain relations present in the data. But model selection must be perceived as an integral part of data analysis (Johnson & Omland, 2004).

Theoretical biology has been focusing largely on dynamics, stochastic processes and discrete mathematics. Nevertheless, neither graphical pathway maps, nor "ODEs or formal programs can indicate whether a description is complete in the sense that it mirrors the full complexity of reality with … its alternative pathways. Even if it predicts a correct observation, we are not to know whether the description is correct, we merely know that it is a possible solution. It is not obvious that there is any way of handling the depths of biological complexity..." (Bard et al., 2013).

Now the time is ripe to rethink the foundations of biological and medical discovery. Our proposal is intended to overcome these shortcomings.

## 8. What do we suggest?

In (Krakauer et al., 2011) the authors plead that it were profitable to explore a wider range of mathematical ideas, such as those connected with logical formalisms, category theory and a variety of frameworks supporting concepts related to information-processing (e.g. info-max assumptions) and forms of distributed decision-making. Hoffman's last article in this issue suggests algebraic geometry (Felix et al., 2008) as a promising theory to tackle the problems of formalizing living systems (Hoffman, 2013). These clarion calls are in line with the Integral Biomathics program (Simeonov, 2010; Simeonov et al., 2012a/b).

As a first practical step towards the realization of this vision, we propose to devise a more moderate, but integrative and evolutionary methodology. This includes the development of a prototype for a unified framework in biomathematics and biocomputation to explore morphogenesis, development, evolution and brain function across multiple domains and perspectives.

---
[15] Multiple Organ Dysfunction Syndrome

This unified research and development framework comprises the following elements:

- Qualitative and quantitative modeling
- Deterministic and stochastic approaches
- Discrete and continuous specifications
- Formal and empirical validation and verification
- Visualization tools supporting both the understanding of the target biological system and the optimization of cognitive exploration processes

Our approach addresses directly three of the 23 DARPA challenges for mathematics[16]:

i) The Mathematics of the Brain,
ii) The Dynamics of Networks, and
iii) The Fundamental Laws of Biology

Six other challenges on this list are also indirectly addressed in our discussion circle. Following previous studies carried out in the INBIOSA project[17] (Simeonov et al., 2011; Simeonov et al. 2012a/b,) and the follow-up activities such as this special issue of JPBMB, we are suggesting an innovative multi-layer approach for qualitative modeling and simulation of living systems based on the integration of three complementary approaches:

- a formal mathematical methodology based on a dynamic category theory, called Memory Evolutive Systems, MES (Ehresmann and Vanbremeersch, 2007).

- a formal virtual computation and communication methodology based on temporal logic, called the Wandering Logic Intelligence, WLI (Simeonov, 2002a/b), and

- a visual modeling and simulation environment GSMP (Siregar et al., 2003).

It represents a formal framework, capable of describing, tracing and predicting the emergence and development of biological structures, using recursive specification and validation in terms of their abstract syntax and operational semantics.

*Most importantly, we are going to provide complementary 1$^{st}$ and 3$^{rd}$ person system descriptions[18] using a Turing Oracle Machine (TOM) implementation in the WLI[19] formalism, thus bridging the gap between the formal and informal/narrative/visual modeling systems[20]*.

Of course, such a system description addressing a challenging question like understanding the movements of viruses within the host organism (Zaichik et al., 2013) also implies a context (Kitto & Kortschak, 2013). The latter needs to be considered from a much broader perspective[21] (Gabora et al., 2013; Gare, 2013), which is of interest for further research.

One *necessary prerequisite* to making fields like neurovirology work under a Turing oracle machine design, is to get *first hand observational data*. Once the data become available, we can evaluate the performance of a TOM at our disposal. Finally, this research framework is intended to interface to third party observation, modeling and data analysis techniques such as used in other quantitative approaches (Wierling et al., 2012).

---

[16] https://www.fbo.gov/spg/ODA/**DARPA**/CMO/BAA07-68/packages.html.
[17] EC FP7 Grant number 269961, www.inbiosa.eu
[18] agent-based simulations could be held to be first person
[19] To our knowledge, there is no current theory, computational framework, or applied field such as system biology in which oracles or meta-level decision rules are used to model living systems in their full complexity.
[20] Applying mathematics is analogical and metaphorical. Mathematical modeling in the future will have to combine both digital and analogical computation, as well as some narrative/visual form of concept expression. Some parts of it will have to be in situ in the medium being investigated, rather than operating in an abstract Platonic world outside the real world we are living in and investigating.
[21] incl. evolution, ecology, society and the researcher engaging in scientific enquiry, theorizing, measuring and communicating results as part of the context being investigated

A first effort to estimate the match and the benefits of unifying these two approaches was made in the WLIMES concept (Ehresmann and Simeonov, 2012). Later, in discussions with other colleagues, we realized that an appropriate integration of structural visualization techniques for representation, discovery and validation/verification, as well as the implementation of modern programming paradigms, into our tailored qualitative approach will be crucial for its success. The major advantage of our approach is that it tries to mimic "the human capacity to combine concepts in new ways or redefine one concept by re-examining it in the context of other concepts" (Van Regenmortel, 2004; Hong, 2013).

## 9. Perspectives

Recently there was some work discussing the chances which biology offers for the creation of new mathematical theorems (Cohen, 2004; Sturmfels, 2007; Clairambault, 2012). The case is no different with the here discussed MES and RB. In view of the capabilities presented by CT and RB, we do not really need a new type of mathematics, but rather appropriate ways to specify the biological problems using CT along the lines of MES and RB that in turn allow the artful selection and integration of appropriate mathematical theories in a workable theoretical framework. Hereby the central issue is the accurate specification of biological problems using insightful and creative techniques while switching between different fields of mathematics and biology to ask for key questions, much as Laplace did with physics and chemistry two centuries ago (Root-Bernstein, 2012a; Hong, 2013). But what if we need a mathematical theory (such as the one about self-emerging objects discussed above) that questions the axiomatic foundations of some recognized mathematical theory within the domain of the life sciences and medicine but can be addressed by CT?

In particular, this was the case with using geometry in physics in Einstein's GRT which required use of a non-Euclidian geometry. Such foundations were known by that time thanks to the works of Gauß, Bolyai, Lobatschewsky and Riemann. But what if we do not have such a base to reflect biological phenomena? Many physical theories were created in simultaneity with developing the mathematics required for representing them. Both disciplines were working hand-in-hand for decades, and the bonds became tighter and tighter with each new theory. Today some scientists have difficulties to classify M-theory (string theory) or gauge theory as a mathematical or a physical theory. It represents a kind of synthetic product of both domains. Now, biology is asked to follow the same way. But where would this lead us?

Gödel's theorems delivered the proof for the necessity to downgrade the paradigm of the axiomatic basis of mathematical completeness (Hilbert's dream). Mathematical biologists (Rosen, 1991, 1999; Louie 2009, 2013) argue that such phenomena as impredicativity (self-referencing definition) are unique for biological systems when compared to physical ones. Fuzziness in biology is another concept not easy to handle with available mathematical tools applied in physics. According to Gödel, CT/MES/RB and WLI are also incomplete theories. What does this imply?

Mathematics is not omnipotent. It needs to be developed and applied accurately in other domains of knowledge (including the humanities and art). In particular, CT is a domain of mathematics (not a 'new mathematics'), which encompasses graphs, groups, posets, etc. and thus allows for some unification of mathematics. INTEGRAL BIOMATHICS should be also an extended mathematical domain, developing and integrating mathematical and computational notions well adapted to the problems of biology. MES and RB were developed in this spirit, but many problems remain, so in particular the relations with computations that we try to model using WLI. These issues are parts of our research schedule.

Concluding all this, the important metaphorical lesson to be learned from Gödel's Incompleteness Theorems is that "nothing works for everything". Thus in the context of mathematical biology and Integral Biomathics, all one can ask for are more suitable kinds of mathematics and computation (than, e.g., Newtonian dynamics and Turing machines[22]) to describe biological phenomena.

But whatever it is (the tools of RB, MES and WLI included), there will always lie something outside its expression powers. There will always be some kind of entity to place in the system description as a placeholder for the unexpected (Simeonov et al., 2012b). This entity (a 'homunculus', a 'Maxwell daemon' or a 'Turing oracle') could probably be captured to some extent when given a specific focus when using the complementary $1^{st}$, $2^{nd}$ and $3^{rd}$ person system description perspectives (Matsuno, 2003) as borders of knowing.

We consider science to be not only concerned simply with producing predictive models, but also and more primordially be involved in *understanding* Nature, as Thom argued in defense of his notion of qualitative mathematics (Thom, 1977, 1994). Nevertheless this observation does not mean that we have to stop doing research here. Each time we have accurately identified a problem and its complementary uncertain part, we will have to find something else, another mathematical or computational formalism, to cross the next bridge when we come close to it, as others did long before us. In sum, all we can say is that we are committed to use CT, MES, RB and WLI as our tools for now. And let's see how far they will take us.

* * *

At the end, I would like to comment on a particular characteristic of autopoiesis (Maturana & Varela, 1980) and self-organization (Kauffman, 1993; Müller & Newman, 2003) in biology, the morphogenesis (D'Arcy Thompson, 1917). Indeed the classical works of Rashevsky (Rashevsky, 1940) and Turing (Turing, 1952) on reaction-diffusion systems, as well as the more recent ones of Nüsslein-Vollhard[23] and Gierer & Meinhardt[24] made a substantial progress for understanding this phenomenon. However, there are still some gaps related to the topological properties of living systems that require special attention within the concept of a "bounded autonomy of levels (BAL)" (Mesarovic and Sreenath, 2006). We can regard them as fractions of the elusive "Turing oracle" machine-representation discussed in the previous section. Particularly there are two aspects that need to be considered in a realistic model about the emergence and development of life reflected in morphogenesis:

First, an organism constantly realizes/computes morphological symmetries throughout its developmental cycle. The external environment/ecosystem tends to interfere with these form-inducing functions, which a healthy and young system repairs again and again to restore its structure and functioning. Simultaneously, internal aging processes evoke degeneration of organs and tissues at its own pace (related to an intrinsic metabolic rate). Both internal and external factors constantly cause symmetries to be broken. In this regard, morphogenesis reflects the capability of a living system to unfold and maintain symmetries, while opposing incompatible changes (that are also the outcome of symmetry breakings) and actively fighting deterioration at the price of using-up energy. Topology is capable of expressing the characteristics of symmetries to be embodied, although it is incapable of delineating the route from the degenerated symmetries back to the symmetry-breaking points. On the other hand, a diffeomorphism[25] is capable of deciphering whatever dynamics is proceeding on a smooth manifold, while being incapable of securing the occurrence of an abrupt symmetry-breaking event. Yet, both topology and differential geometry are incapable of dealing with the task of perpetuating an inexhaustible degeneracy of symmetries to be broken.

---

[22] Note: Using a Turing Machine (TM) as a metaphorical model for biology is inappropriate, but using Turing's notion of a Universal Machine (TUM) to define a mathematical biology, which is not based on the TM design, but on Church's thesis is a reasonable approach (Marchal, 1992).
[23] http://www.eb.tuebingen.mpg.de/nc/research/departments/details/details/cnv.html#=1.
[24] http://www.eb.tuebingen.mpg.de/research/emeriti/hans-meinhardt/pattern7.html.
[25] an isomorphism in the category of smooth manifolds, which is an invertible function that maps one differentiable manifold one-to-one-and-onto another, such that both the function and its inverse are smooth (Wikipedia)

Therefore, what could be further developed to handle symmetry breaking of degenerative processes from topological perspective within MES (Ehresmann and Vanbremeersch, 2007) is the inclusion of concepts rooted possibly in the "laws of form" calculus (Spencer-Brown, 1972) and incorporating perhaps concepts from differential geometry (Kühnel, 2005), group and symmetry theories (Carter, 1997; McWeeny, 2012), knot and braid theories (Moran, 2000; Kauffman, 2006), Temperley-Lieb algebra (Abramsky, 2008), as well as quantum topology and computation (Kauffman & Baadhio, 1993; Pachos, 2012), even supersymmetry (LaBelle, 2005) and beyond, all within a unifying dynamic category-theoretical-computational framework such as the one discussed in this paper.

Second, what is required for delivering a realistic model of developing life forms is the construction of a material body as a computational agent inside those hypothetical "protogerms" that is able to maintain its own identity while constantly suffering a breaking of the symmetries latent in itself. This material body "keeping its self-identity upon variability"[26] can be regarded as the primitive (bio)computational element or engine: the *vir*[27]. The living cell itself, or its ingredient macromolecules (including DNA/RNA complexes) are, in fact, the resulting computational assemblies – the "clusters" and "clouds" built from the operation of such multiple primitive machines (viri). On the other hand, premature aging, infections (such as those caused by viruses, fungi and bacteria) and degenerative/metabolic disorders can be seen as biological entities with (distributed) computations performed by exo-viri[28] provoking structural and functional organic changes in the opposite direction towards deterioration, deformation, dysfunction and death. Both analogous biocomputational mechanisms, the ones of (eso-)viri and of exo-viri, concur inside the organism and in the ecosystem as a whole. Nevertheless, *senescence* entails a natural, inevitable end of life (Salthe, 2013); see, however, also (Kloeden, Rössler and Rössler, 1990).

Finally, the key issue remaining to be discovered and formalized in morphogenesis is how *decision making* in these opposite biocomputational processes and senescence is implemented, in contrast to the scheme suggested e.g. by Gierer & Meinhardt which (according to some interpreters) imposes an arbitrary boundary condition of hypothetical origin. It is one thing to consider the role of topology in morphogenesis, but quite another to see what aspect of the system is responsible for making morphogenetic decisions. This issue could be resolved using topology and geometry backed e.g. by category theory and algebraic geometry, and some sort of biocomputational logic such as those discussed in (Simeonov, 2002a/b) and (Goranson & Cardier, 2013) explaining the dynamics of the underlying processes.

Of course, all of this goes without saying that understanding the dynamics of biology does require something *invariant* in its own descriptive enterprise, just as the dynamics of physics does. What is relevant exclusively for biology is the **occurrence of a function, which keeps form invariant** throughout the exchange of the component materials (Matsuno, 2013).

---

**Acknowledgements**: I would like to thank Koichiro Matsuno, Stanley N. Salthe, Otto E. Rössler, Arran Gare, Andree C. Ehresmann, Jaime Gomez-Ramirez, Pridi Siregar, Aloisus H. Louie, Bruno Marchal, Ted Goranson, Louis H. Kauffman and Robert S. Root-Bernstein for their valuable help with reviewing and commenting on this paper. It has passed multiple peer reviews and reformulations (from individual 1[st], through shared 2[nd] into consolidated 3[rd] person descriptions) to hopefully offer a semi-formal proposal implementing the requirements of a progress report in the field.

---

[26] The object addressable in 3[rd] person description is actually being upheld in 1[st] person description (Koichiro Matsuno, June 12[th] 2013, personal correspondence).
[27] man, hero, man of courage (Lat.); this element does not exclude the possible concomitance and operation of a "Turing oracle" machine within itself.
[28] A pathogen such as a virus infecting an organism can incorporate such an *exo-vir* body.

# References

Abramsky, S. 2008. Temperley-Lieb Algebra: From Knot Theory to Logic and Computation via Quantum Mechanics. In: *Mathematics of Quantum Computing and Technology*, G. Chen, L. Kauffman and S. Lomonaco (Eds.). Taylor and Francis, 415-458, 2008. http://arxiv.org/pdf/0910.2737v1.pdf.

Aderem A, Smith K.D. 2004. A systems approach to dissecting immunity and inflammation. *Semin. Immunol.* 16: 55–67. http://www.ncbi.nlm.nih.gov/pubmed/14751764.

Aerts, D., Gabora, L. 2005. A state-context-property model of concepts and their combinations I/II. *Kybernetes, 34*(1&2), 167-191; 192-221. (Special issue dedicated to Heinz Von Foerster.)

Ahn, A. C., Tewari, M., Poon, C.-S., Phillips, R. S. 2006. The limits of reductionism in medicine: could systems biology offer an alternative? *PLoS medicine*, 3(6):e208.

Alm E., Arkin A.P. 2003. Biological networks. *Curr. Opin. Struct. Biol.* **13**: 193–202. http://www.ncbi.nlm.nih.gov/pubmed/12727512.

Bard, J., Melham, T., Noble, D., 2013. Editorial. Epilogue: Some conceptual foundations of systems biology. *J. Progress in Biophysics and Molecular Biology* XXX (2013) 1–3. http://dx.doi.org/10.1016/j.pbiomolbio.2012.12.002.

Brenner, S. 2010. Sequences and consequences. *Phil. Trans. R. Soc. B.* 12 January 2010 vol. 365 no. 1537, 207-212.

Camazine, S., Deneubourg, J.-L., Franks, N., Sneyd, J., Therula, G., Bonabeau, E., *Self-organisation in Biological Systems*. 2003. Princeton University Press. ISBN-10: 0691116245; ISBN-13: 978-0691116242.

Cardinale, S., Arkin, A.P. 2012. Contextualizing context for synthetic biology--identifying causes of failure of synthetic biological systems. *Biotechnol. J.* 7(7):856-66. doi: 10.1002/biot.201200085. http://www.ncbi.nlm.nih.gov/pubmed/22649052.

Carter. L. R. 1997. *Molecular Symmetry and Group Theory*. Wiley. ISBN-10: 0471149551; ISBN-13: 978-0471149552.

Cazalis, R. 2013. The organism and its double: An approach to the individual. *Journal Progress in Biophysics and Molecular Biology.* Special Issue: "Can biology create a profoundly new mathematics?" Available online 27 March 2013. http://dx.doi.org/10.1016/j.pbiomolbio.2013.03.006. In this issue.

Clairambault, J. 2012. Can theorems help treat cancer? *Journal of Mathematical Biology.* DOI: 10.1007/s00285-012-0518-9**.** http://hal.archives-ouvertes.fr/docs/00/78/02/03/PDF/JC4JMBCantheorems2012Preprint.pdf**.**

Cohen, J. E. 2004. Mathematics is biology's next microscope, only better; biology is mathematics' next physics, only better, *PLOS Biology* **2** (2004) No.12.

Cook, S. 1971. The complexity of theorem proving procedures. *Proceedings of STOC'71, The Third Annual ACM Symposium on Theory of Computing*. 151–158. ACM Press.

Csete ME, Doyle JC (2002) Reserve engineering of biological complexity. Science 295: 1664–1669. http://www.ncbi.nlm.nih.gov/pubmed/11872830.

Dennet, D. 1988.Quining Qualia. In: A. Marcel and E. Bisiach, (Eds), *Consciousness in Modern Science*, Oxford University Press 1988. Reprinted in W. Lycan, (Ed.), *Mind and Cognition: A Reader*, MIT Press, 1990; A. Goldman, (Ed.) *Readings in Philosophy and Cognitive Science*, MIT Press, 1993. http://ase.tufts.edu/cogstud/papers/quinqual.htm. Retrieved 2013-06-05.

Drews, J. 2003. Strategic trends in the drug industry. Drug Discov. Today 8: 411–420. http://www.ncbi.nlm.nih.gov/pubmed/12706659.

Edelman, G. M., Gally, J. A. 2001. Degeneracy and complexity in biological systems. *PNAS*. vol. 98 no. 24, 13763–13768. http://www.pnas.org/content/98/24/13763.long.

Ehresmann, A., Vanbremeersch, J.-P. 2007, *Memory Evolutive Systems:* Hierarchy, Emergence, Cognition, Elsevier. ISBN-10: 0444522441; ISBN-13: 978-0444522443.

Ehresmann, A., 2012. MENS, an info-computational model for (neuro-)cognitive systems up to creativity, *Entropy* 14, 2012, 1703-1716 (Open Access).

Ehresmann, A. C., Simeonov, P. L. 2012. WLIMES: Towards a Theoretical Framework for Wandering Logic Intelligence Memory Evolutive Systems. In: *Integral Biomathics: Tracing the Road to Reality, Proceedings of iBioMath 2011, Paris and ACIB '11, Stirling UK,* P. L. Simeonov, L. S. Smith, A. C. Ehresmann (Eds.), Springer-Verlag, ISBN-10: 3642281109; ISBN-13: 978-3642281105.

Felix, Y., Oprea, J., Tanre, D., 2008. *Algebraic Models in Geometry*. Oxford Univ. Press, New York. ISBN-10: 019920652X; ISBN-13: 978-0199206520.

Gabora, L., Scott, E. O., Kauffman, S. 2013. A quantum model of exaptation: Incorporating potentiality into evolutionary theory. *Journal Progress in Biophysics and Molecular Biology.* Special Issue: "Can biology create a profoundly new mathematics?" Available online 06 April 2013. ISSN: 0079-6107. http://dx.doi.org/10.1016/j.pbiomolbio.2013.03.012. In this issue.

Gare, A. 2013. Overcoming the Newtonian paradigm: The unfinished project of theoretical biology from a Schellingian perspective. *Journal Progress in Biophysics and Molecular Biology.* Special Issue: "Can biology create a profoundly new mathematics?" Available online 06 April 2013. ISSN: 0079-6107. In this issue.

Glassman R.H., Sun A.Y. 2004. Biotechnology: identifying advances from the hype. *Nat. Rev. Drug Discov*. 3: 177–183. http://www.ncbi.nlm.nih.gov/pubmed/15040581.

Gomez-Ramirez, J., Wu. J. 2012. *A new vision for biomedicine: A systems approach*. In: *2012 ICME International Conference on Complex Medical Engineering (CME),* 479–484.

Gomez-Ramirez, J. , Sanz, R. 2013. On the limitations of standard statistical modeling in biological systems: a full Bayesian approach for biology. *J. Progress in Biophysics and Molecular Biology*. Special Theme Issue on Integral Biomathics: Can Biology Create a Profoundly New Mathematics and Computation? Elsevier. ISSN: 0079-6107. In this issue.

Goranson, T., Cardier, B. 2013. A Two-sorted Logic for Structurally Modeling Systems. *J. Progress in Biophysics and Molecular Biology. Special Theme Issue on Integral Biomathics: Can Biology Create a Profoundly New Mathematics and Computation?* Elsevier. ISSN: 0079-6107. http://dx.doi.org/10.1016/j.pbiomolbio.2013.03.015, In this issue.

Heiner, M., Gilbert, D. 2013. From Petri nets to partial differential equations and beyond. BioModel Engineering for Multiscale Systems Biology. *J. Progress in Biophysics and Molecular Biology.* Available Online 12 October 2012. http://dx.doi.org/10.1016/j.pbiomolbio.2012.10.001.

Hoffman, W C., 2012. The dialectics of mind. *Journal of Mind Theory* (Madrid) **1 (**1), 1 – 24. Electronic: http://www.aslab.upm.es/documents/journals/JMT/Vol1-No1/JMT_1_1-DIA-HOFFMAN.pdf.

Hoffman, W. C. 2013. Mathematics for Biomathics. J*. Progress in Biophysics and Molecular Biology*. http://dx.doi.org/10.1016/j.pbiomolbio.2013.03.016. In this issue.

Hong, F. T. 2013. The Role of Pattern Recognition in Creative Problem Solving: A Case Study in Search of New Mathematics for Biology. *J. Progress in Biophysics and Molecular Biology. Special Theme Issue on Integral Biomathics: Can Biology Create a Profoundly New Mathematics and Computation?* Elsevier. ISSN: 0079-6107. In this issue.

Johnson, J. B., Omland, K .S. 2004. Model selection in ecology and evolution. *Trends in ecology & evolution*, 19(2):101–108, February, 2004. PMID: 16701236.

Junier, I., Martin, O., Képès, F. 2010. Spatial and topological organization of DNA chains induced by gene co-localization. *PLoS Comput Biol*. 6(2):e1000678. doi: 10.1371/journal.pcbi.1000678. http://gennetec.csregistry.org/tiki-download_wiki_attachment.php?attId=226.

Kauffman, L. H., Baadhio, R. A. 1993. *Quantum Topology.* Series on Knots. World Scientific. ISBN-10: 9810215444; ISBN-13: 978-9810215446.

Kauffman, L. H. 2001. The Mathematics of Charles Sanders Peirce. *Cybernetics & Human Knowing*, Vol. 8, No.1–2, 79-110. http://homepages.math.uic.edu/~kauffman/CHK.pdf.

Kauffman, L. H. 2006. *Formal Knot Theory*. Dover. ISBN-10: 048645052X; ISBN-13: 978 0486450520.

Kauffman, S. A. 1993. *The Origins of Order: Self-Organization and Selection*. Oxford University Press. ISBN-10: 0195079515; ISBN-13: 978-0195079517.

Kim, J. 1984. Concepts of Supervenience. *Philosophy and Phenomenological Research* ,45, 2: 153-176.

Kim, J. 1987. 'Strong' and 'Global' Supervenience Revisited. *Philosophy and Phenomenological Research,* 48, 2: 315-326.

Kim J. 1999. Making sense of emergence. *Philos. Stud.* 95: 3–36.

Kitano, H. 2002 Systems biology: a brief overview. *Science*. 295: 1662–1664. http://www.ncbi.nlm.nih.gov/pubmed/11872829.

Kitto, K., Kortschak, R. D., 2013. Contextual models and the non-Newtonian paradigm. *J. Progress in Biophysics and Molecular Biology*. Special Theme Issue on Integral Biomathics: Can Biology Create a Profoundly New Mathematics and Computation? Elsevier. ISSN: 0079-6107. In this issue.

Kloeden P. E., Rössler, O. E., Rössler, R**.** 1990. A predictive model for life expectancy curves. *Biosystems* 24, 119-125**.**

Kolodkin, A., Simeonidis, E., Westerhoff, H. V. 2012. Computing life: Add logos to biology and bios to physics. *J. Progress of Biophysics and Molecular Biology* . Available online 24 October 2012. http://dx.doi.org/10.1016/j.pbiomolbio.2012.10.003.

Krakauer, D. C., Collins, J. P., Erwin, D., Flack, J. S., Fontana, W., Laubichler, M. D., Prohaska, S. J., West, J. B., Stadler, P. F. 2011. The challenges and scope of theoretical biology. 50th Anniversary Year Review. *Journal of Theoretical Biology* 276 (2011) 269-276. doi:10.1016/jjtbi.2011.01.051.

Kubinyi, H. 2003. Drug research: myths, hype and reality. *Nat. Rev. Drug Discov.* 2: 665–668. http://www.ncbi.nlm.nih.gov/pubmed/12904816.

Kühnel, W. 2005. *Differential Geometry: Curves –Surfaces – Manifolds*. American Mathematical Society. 2nd Edition. ISBN-10: 0821839888; ISBN-13: 978-0821839881.

LaBelle, P. 2005. *Supersymmetry DeMYSTiFied*. McGraw-Hill. ASIN: B009YXFX0G.

Lewontin, R. 2002. *The Triple Helix. Gene, Organism and Environment*. Harvard University Press, Cambridge, MA, USA. ISBN-10: 0674006771; ISBN-13: 978-0674006775.

Louie, A. H. 2009. *More than Life Itself: A Synthetic Continuation in Relational Biology*. Ontos. ISBN-10: 3868380442; ISBN-13: 978-3868380446.

Louie, A. H. 2013. *The Reflection of Life: Functional Entailment and Imminence in Relational Biology* (IFSR International Series on Systems Science and Engineering). Springer. ISBN-10: 1461469279; ISBN-13: 978-1461469278.

Mac Lane, S. 1998. *Categories for the Working Mathematician*. Graduate Texts in Mathematics 5 (2nd ed.). Springer-Verlag. ISBN-10: 0387984038; ISBN-13: 978-0387984032.

Marchal, B. 1992. Amoeba, Planaria, and Dreaming Machines. In: P. Bourgine and F. J. Varela (Eds.), *Artificial Life, Towards a Practice of Autonomous Systems*, Proc. ECAL 91, MIT Press, 429-440.

Marchal, B. 2013. The Computationalist Reformulation of the Mind-Body Problem. *J. Progress in Biophysics and Molecular Biology. Special Theme Issue on Integral Biomathics: Can Biology Create a Profoundly New Mathematics and Computation?* Elsevier. ISSN: 0079-6107. In this issue.

Matsuno, K. 1989. *Protobiology: Physical Basis of Biology*. Boca Raton, FL: CRC Press. ISBN-10: 0849364035; ISBN-13: 978-0849364037.

Matsuno, K. 1996. Internalist Stance and the Physics of Information. *BioSystems* **38**, 111–118.

Matsuno, K. 2003. Quantum Mechanics in First, Second and Third Person Descriptions. *BioSystems* **68** (2-3):107-18.

Matsuno, K., 2013. Toward Accommodating Biosemiotics with Experimental Sciences. *Biosemiotics* 6, 125-141.

Maturana, H., Varela, F., 1980. *Autopoiesis and Cognition. The Realization of the Living. D. Reidel Pub. Co., Holland*. ASIN: B002HS1I8Q. Also in: Springer-Verlag *(August 31, 1991)*. ISBN-10: 9027710163; ISBN-13: 978-9027710161.

McWeeny, R. 2012. *Symmetry: An Introduction to Group Theory and Its Applications*. Dover. ASIN: B008TVLKEC.

Mesarovic, M.D., Sreenath, S.N., 2006. Beyond the flat earth perspective in systems biology. *Biol. Theory* 1 (1), 33-34.

Moran, S. 2000. The Mathematical Theory of Knots and Braids: An Introduction. North Holland. ISBN-10: 0444557741; ISBN-13: 978-0444557742.

Morowitz H.J. (2002) *The Emergence of Everything. How the World Became Complex*. Oxford University Press, ISBN-10: 0195173317; ISBN-13: 978-0195173314.

Mossio, M., Moreno, A. 2010. Organisational Closure in Biological Organisms. *Hist. Phil. Life Sci.* 32. 269-288. http://mossio.free.fr/Publications_files/Pdf/Mossio_Moreno_hpls_2010.pdf

Müller, G. B., Newman, S. 2003. *Origination of Organismal Form: Beyond the Gene in Developmental and Evolutionary Biology*. A Bradford Book. ISBN-10: 0262134195; ISBN-13: 978-0262134194.

Nakajima, T. 2013. Probability in biology: overview of a comprehensive theory of probability in living systems. *J. Progress in Biophysics and Molecular Biology*. Special Theme Issue on Integral Biomathics: Can Biology Create a Profoundly New Mathematics and Computation? Elsevier. ISSN: 0079-6107. In this issue.

Noble, D. 2008. *The Music of Life: Biology Beyond Genes*. Oxford University Press. ISBN-10: 0199228361; ISBN-13: 978-0199228362.

Noble, D. 2012. A theory of biological relativity: no privileged level of causation. *Interface Focus* RS. February 6, 2012 2, 1, 55-64.

Noble, D. 2013. A biological relativity view of the relationships between genomes and phenotypes. J. Prog. Biophys. Mol. Biol. Available online 5 October 2012. http://dx.doi.org/10.1016/j.pbiomolbio.2012.09.004.

Omholt, S., 2013. From sequence to consequence and back. *J. Prog. Biophys. Mol. Biol.* Available online 25 September 2012. http://dx.doi.org/10.1016/j.pbiomolbio.2012.09.003.

Pattee, H. H. 1973. *Hierarchy Theory. The Challenge of Complex Systems.* George Braziller. ISBN-10: 080760674X. ISBN-13: 978-0807606742.

Pachos, J., K. 2012. *Introduction to Topological Quantum Computation*. Cambridge University Press. ISBN-10: 1107005043; ISBN-13: 978-1107005044.

Rashevsky. N. 1940. An approach to the mathematical biophysics of biological self-regulation and the cell polarity. *Bull. Math. Biophys*. 2, 15-25.

Root-Bernstein, R. S., Dillon, P. F. 1997. Molecular complementarity, I: The molecular complementarity theory of the origin and evolution of life. *J. Theor. Biol*. 188, 447–479.

Root-Bernstein, R. S. 2012a. Processes and Problems That May Define the New BioMathematics Field. In: *Integral Biomathics: Tracing the Road to Reality*, Proc. of iBioMath 2011, Paris and ACIB '11, Stirling UK, P. L. Simeonov, L. S. Smith, A. C. Ehresmann (Eds.), Springer-Verlag, Heidelberg, ISBN-10: 3642281109; ISBN-13: 978-3642281105.

Root-Bernstein, R.S. 2012b. A modular hierarchy-based theory of the chemical origins of life, based on molecular complementarity. *Accounts Chem. Res*. 45(12):2169-77. doi: 10.1021/ar200209k. http://www.ncbi.nlm.nih.gov/pubmed/22369101.

Rosen, R., 1958a. A relational theory of biological systems. Bull. Math. Biophys. 20, 245-260. Springer-Verlag. ISSN: 0092-8240 (Print) 1522-9602 (Online).

Rosen, R., 1958b. The representation of biological systems from the standpoint of the theory of categories. Bull. Math. Biophys. 20, 317e341. Springer-Verlag. ISSN: 0092-8240 (Print) 1522-9602 (Online).

Rosen, R., 1959. A relational theory of biological systems II. Bull. Math. Biophys. 21,109-128. Springer-Verlag. ISSN: 0092-8240 (Print) 1522-9602 (Online).

Rosen, R., 1991. *Life Itself.* Columbia University Press, New York, ISBN 0-231-07565-0.

Rosen, R., 1999. *Essays on Life Itself.* Columbia University Press, New York, ISBN 0-231-10510-X.

Rosen, R. 2012. *Anticipatory Systems: Philosophical, Mathematical, and Methodological Foundations (International Federation for Systems Research International Series on Systems Science and Engineering, Vol. 1)*. Springer; 2nd edition (February 2, 2012). ISBN-10: 1461412684; ISBN-13: 978-1461412687.

Rössler, O. E., 1987. Endophysics. In: J. L. Casti, A. Karlqvist (Eds). *Real Brains, Artificial Minds*. Elsevier. ISBN-10: 0444011552; ISBN-13: 978-0444011558 25-46.

Rössler, O.E., 1998. *Endophysics: the World as an Interface.* World Scientific, ISBN 981-02-2752-3.

Salthe, S. N. 1985. *Evolving Hierarchical Systems*. Columbia University Press. ISBN-10: 0231060165; ISBN-13: 978-0231060165.

Salthe, S. N. 2008. Natural selection in relation to complexity. *Artificial Life* 14 (3): 363-374.

Salthe, S. N. 2012. Hierarchical structures. *Axiomathes.* 22: 355-383.

Salthe, S. N. 2013. To naturally compute (something like) biology. *Journal Progress in Biophysics and Molecular Biology.* Special Issue: "Can biology create a profoundly new mathematics?" http://dx.doi.org/10.1016/j.pbiomolbio.2013.03.005. In this issue.

Salthe, S. N., Matsuno, K., 1995. Self-organization in Hierarchical Systems. *J. Soc. Evol. Syst.* **18**, 327-338.

Schroeder, M. J. 2013. Crisis in science: In search for new theoretical foundations. *Journal Progress in Biophysics and Molecular Biology.* Special Issue: "Can biology create a profoundly new mathematics?" Available online 27 March 2013. http://dx.doi.org/10.1016/j.pbiomolbio.2013.03.003. In this issue.

Shapiro, J. A. 2013. Rethinking the (im)possible in evolution. *Journal Progress in Biophysics and Molecular Biology*. Volume 111, Issues 2–3, April 2013, 92–96. Elsevier. http://www.sciencedirect.com/science/article/pii/S0079610712000909.

Siegelmann, H. 2013. Turing on Super-Turing and Adaptivity. *J. Progress in Biophysics and Molecular Biology. Special Theme Issue on Integral Biomathics: Can Biology Create a Profoundly New Mathematics and Computation?* Elsevier. ISSN: 0079-6107. In this issue.

Simeonov, P. L. 2002a. The Viator Approach: About Four Principles of Autopoietic Growth On the Way to Hyperactive Network Architectures*, Annual IEEE Workshop on Fault-Tolerant Parallel and Distributed Systems (FTPDS'02) | 2002 IEEE International Symposium on Parallel & Distributed Processing (IPDPS'02)*, April 15-19, 2002, Ft. Lauderdale, FL, USA, IEEE Computer Society, Washington, DC, USA, pp. 320 - 327, ISBN:0-7695-1573-8, http://ieeexplore.ieee.org/iel5/7926/21854/01016528.pdf.

Simeonov, P. L., 2002b. *The Wandering Logic Intelligence, A Hyperactive Approach to Network Evolution and Its Application to Adaptive Mobile Multimedia Communications*. Ph. D. Thesis. 2002. Technische Universität Ilmenau, Fakultät für Informatik und Automatisierung, http://deposit.ddb.de/cgi-bin/dokserv?idn=974936766.

Simeonov, P. L. 2002c. WARAAN: A Higher-Order Adaptive Routing Algorithm for Wireless Multimedia in Wandering Networks*, 5th IEEE International Symposium on Wireless Personal Multimedia Communications (WPMC'2002)***,** October 27-30, 2002, Honolulu, Hawaii, USA, 1385-1389, http://ieeexplore.ieee.org/iel5/8154/23649/01088407.pdf.

Simeonov, P. L., Ehresmann, A. C., Smith, L. S., Gomez-Ramirez, J., Repa, V. 2011. A New Biology: A Modern Perspective on the Challenge of Closing the Gap between the Islands of Knowledge. In: Cezon, M., Wolfsthal, Y. (eds.) *ServiceWave 2010 Workshops (EDBPM 2010)*, Ghent, Belgium, December 2010. LNCS 6569, Springer-Verlag, Heidelberg, 2011, 188-195.

Simeonov, P. L., Smith, L., S., Ehresmann, A. C. (Eds.), 2012a. *Integral Biomathics: Tracing the Road to Reality*, Proceedings of iBioMath 2011, Paris and ACIB '11, Stirling UK. Springer-Verlag, Heidelberg. ISBN-10: 3642281109; ISBN-13: 978-3642281105.

Simeonov, P. L., Brezina, E., Cottam, Ehresmann, A. C., Gare, A., Goranson, T., Gomez-Ramirez, J., Josephson, B. D., Marchal, B., Matsuno, K., Root-Bernstein, R. S., Rössler, O. E., Salthe, Schroeder, M., S. N., Seaman, Siregar, P., B., Smith, L. S. 2012b. Stepping Beyond the Newtonian Paradigm in Biology. Towards an Integrable Computational Model of Life: Accelerating Discovery in the Biological Foundations of Science. INBIOSA White Paper. In: *Integral Biomathics: Tracing the Road to Reality*, Proc. of iBioMath 2011, Paris and ACIB '11, Stirling UK, P. L. Simeonov, L. S. Smith, A. C. Ehresmann (Eds.), Springer-Verlag, Heidelberg, ISBN-10: 3642281109; ISBN-13: 978-3642281105.

Siregar, P., Julien, N., Sinteff, J. P. 2003. Computational Integrative Physiology: At the Convergence of the Life, Physical and Computational Sciences, *Methods of Information in Medicine,* 42:177-84, 2003.

Spencer-Brown, G. 1972. *Laws of Form*. Crown Publishers, ISBN 0-517-52776-6.

Sturmfels, B. 2007. *Can Biology Lead to New Theorems?* Talk. http://math.berkeley.edu/~bernd/ClayBiology.pdf.

Thom, R. 1977. Structural Stability, Catastrophe Theory and Applied Mathematics. *SIAM Review*, Vol. 19, No. 2. April 1977. 189-201.

Thom, R. 1994. *Structural Stability And Morphogenesis*. Westview Press. ISBN-10: 0201406853; ISBN-13: 978-0201406856.

Thompson, D. W., 1917. *On Growth and Form.* Cambridge Univ. Press, ISBN 0-521-43776-8.

Tschernyschkow et al., 2013. Rule-based modeling and simulations of the inner kinetochore structure. *Journal Progress in Biophysics and Molecular Biology.* Special Issue: "Can biology create a profoundly new mathematics?" Available online 02 April 2013. http://dx.doi.org/10.1016/j.pbiomolbio.2013.03.010. In this issue.

Turing, A. 1939. Systems of Logic based on Ordinals. *Proc. London Math. Soc.* Ser. 2 45, 158–226. http://www.turingarchive.org/browse.php/B/15.

Turing, A. 1952. The chemical basis of morphogenesis. *Philosophical Transactions of the Royal Society of London*. Series B, Biological Sciences, Vol. 237, No. 641. (Aug. 14, 1952), 37-72. http://www.dna.caltech.edu/courses/cs191/paperscs191/turing.pdf.

Weng, G., Bhalla, U. S., Iyengar, R. 1999 Complexity in biological signalling systems. *Science* 284: 92–96. http://www.ncbi.nlm.nih.gov/pubmed/10102825

Van Regenmortel, M. H. V., 2004. Reductionism and complexity in molecular biology. *EMBO Rep*. 2004 November; 5(11): 1016–1020. doi: 10.1038/sj.embor.7400284.

Vrobel, S. 2013. Measuring the temporal extension of the Now. *Journal Progress in Biophysics and Molecular Biology.* Special Issue: "Can biology create a profoundly new mathematics?" http://dx.doi.org/10.1016/j.pbiomolbio.2013.03.009. In this issue.

Wepiwé, G., Simeonov, P. L. 2005. A Concentric Multi-ring Overlay for Highly Reliable P2P Networks. *4th IEEE International Symposium on Network Computing and Applications (NCA'05)*. July 27-29, 2005, Cambridge, Massachusetts, USA, 83-90. http://doi.ieeecomputersociety.org/10.1109/NCA.2005.1

Werner, E., 2005. Genome semantics, in silico multicellular systems and the central dogma. FEBBS Lett. 579, 179–182.

Wierling, C., Kühn , A., Hache , H., Daskalaki, A., Maschke-Dutz, E., Peycheva, S., Li, J., Herwig, R., Lehrach, H. 2012. Prediction in the face of uncertainty: A Monte Carlo-based approach for systems biology of cancer treatment. Elsevier. J. Mutation Research. 746 (2012) 163– 170.

Williams, D.A., Baum, C. 2003. Gene therapy: new challenges ahead. *Science* 302: 400–401. http://www.ncbi.nlm.nih.gov/pubmed/14563994.

Yoshiteru, I. 2009. A Note on Biological Closure and Openness: A System Reliability View. *Proc. of KES'09 Knowledge-Based and Intelligent Information and Engineering Systems.* Springer. Lecture Notes in Computer Science. Vol. 5712, 805-812.

Zadeh, L. A., 2000. *Fuzzy Sets and Fuzzy Information Granulation Theory*. Key Selected Papers by Lotfi A. Zadeh. Beijing Normal University Press. ISBN-10: 7303053247; ISBN-13: 978-7303053247.

Zaichick, S. V., Bohannon, K. P., Hughes, A., Sollars, P. J., Pickard, G. E., Smith, G. A. The Herpesvirus VP1/2 Protein Is an Effector of Dynein-Mediated Capsid Transport and Neuroinvasion. *Cell Host & Microbe*, 2013; 13 (2): 193 DOI: 10.1016/j.chom.2013.01.009.